# Refuting the Pseudo Attack on the REESSE1+ Cryptosystem[*]

Shenghui Su [1, 2], and Shuwang Lü [3]

[1] College of Computer Science, Beijing University of Technology, Beijing 100022, P.R.China

[2] School of Info Engi, University of Science & Technology Beijing, Beijing 100083, P.R China

*sheenway@126.com*

[3] School of Graduate, Chinese Academy of Sciences, Beijing 100039, P.R.China

*swlu@ustc.edu.cn*

**Abstract**: We illustrate through example 1 and 2 that the condition at theorem 1 in [8] dissatisfies necessity, and the converse proposition of *fact* 1.1 in [8] does not hold, namely the condition $Z/M - L/A_k < 1/(2A_k^2)$ is not sufficient for $f(i)+f(j)=f(k)$. Illuminate through an analysis and ex.3 that there is a logic error during deduction of *fact* 1.2, which causes each of *fact* 1.2, 1.3, 4 to be invalid. Demonstrate through ex.4 and 5 that each or the combination of $q_{u+1} > q_u \Delta$ at *fact* 4 and *table* 1 at *fact* 2.2 is not sufficient for $f(i)+f(j)=f(k)$, *property* 1, 2, 3, 4, 5 each are invalid, and *alg*.1 based on *fact* 4 and *alg*.2 based on *table* 1 are disordered and wrong logically. Further, manifest through a repeated experiment and ex.5 that the data at *table* 2 is falsified, and the example in [8] is woven elaborately. We explain why $C_x \equiv A_x W^{f(x)}$ (% $M$) is changed to $C_x \equiv (A_x W^{f(x)})^\delta$ (% $M$) in REESSE1+ v2.1. To the signature fraud, we point out that [8] misunderstands the existence of $T^{-1}$ and $Q^{-1}$ % ($M - 1$), and forging of $Q$ can be easily avoided through moving $H$. Therefore, the conclusion of [8] that REESSE1+ is not secure at all (which connotes that [8] can extract a related private key from any public key in REESSE1+) is fully incorrect, and as long as the parameter $\Omega$ is fitly selected, REESSE1+ with $C_x \equiv A_x W^{f(x)}$ (% $M$) is secure.

## 1  Introduction

In April 2001, we put forward the REESSE1 public-key encryption scheme [1]. In September 2003, we proposed the REESSE1 public-key cryptosystem which is an extension of the first version, and includes both encryption and signature [2]. In May 2005, it was argued that the lever function $\ell(.)$ is necessary and sufficient for the security of the REESSE1 encryption [3]. In [3], the continued fraction method of analyzing the key transforms $C_x \equiv A_x W$ and $C_x \equiv A_x W^{\ell(x)}$ (% $M$) with $x \in [1, n]$ and $\ell(x) \in \Omega$ was mentioned earlier than in any other publications. In November 2006, an abbreviation of the REESSE1+ cryptosystem was submitted to eprint.iacr.org [4].

As is pointed out in [4], the set $\Omega = \{5\delta, …, (n + 4)\delta \mid \delta \geq 1\}$ is not unique, and other $\Omega$ may be selected — $\Omega = \{n + 1, …, n + n\}$ with $\ell(i) + \ell(j) \neq \ell(k) \ \forall \ i, j, k \in [1, n]$ for example. Clearly, $\Omega$ is a security dominant parameter, and just like $p$ and $q$ in the RSA cryptosystem.

In May 2005, [5] pointed out that the REESSE1 signature scheme was insecure, which is right.

In July 2005, [6] thought unreasoningly that the REESSE1 encryption scheme was insecure, which is wrong, and rebutted thoroughly by us in [7]. Moreover, [7] illuminated definitely that the idea of the continued fraction analysis of REESSE1 did not originate from [6] (naturally also not from [8]), ant the idea firstly formally appeared in our 2004 application for a national fund project [7]. What needs to be pointed out further is that the authors of [8] are the related reviewers of our 2004 application.

In December 2006, [8] thought unreasoningly again that the REESSE1+ public-key cryptosystem is not secure at all, which connotes any private key in REESSE1+ can be extracted by [8]. It is of flubdub and gulf.

The ancients said 'stop an advancing army with troops, and stop onrushing water with earth'.

In what follows, the function $f$ in [8] is namely the function $\ell$ in [4], namely $f(i)$, $f(j)$, $f(k)$ in [8] are equivalent to $\ell(i)$, $\ell(j)$, $\ell(k)$, unless otherwise specified, the sign $\bar{M}$ represents '$M-1$', the sign % does 'modulo', and unattached ($x$) does $x$-th expression.

In short, there exist 6 grave faults in [8]:

① The converse proposition of *fact* 1.1 does not hold.

Clearly, *fact* 1.1 implies that if $f(i)+f(j)=f(k)$, then $Z/M - p_u/q_u < 1/(2q_u^2)$ with $L/A_k = p_u/q_u$.

We will prove by a counterexample that the former is only sufficient, but not necessary, namely if $Z/M - p_u/q_u < 1/(2q_u^2)$, then $f(i)+f(j)=f(k)$ do not necessarily hold, and also namely $Z/M - p_u/q_u < 1/(2q_u^2)$ for $f(i)+f(j)=f(k)$ is only necessary, but not sufficient.

② *Fact* 1.2, 1.3 and 4 do not always hold.

Even if they hold, *fact* 1.2, 1.3 and 4 each are insufficient for $f(i)+f(j)=f(k)$, and further, property 1, 2, 3, 4 and 5 are invalid. Note *fact* 4 is essentially equivalent to each of *fact* 1.2 and 1.3.

③ The converse proposition of *fact* 2.2 does not hold, namely *table* 1 is insufficient for $f(i)+f(j)=f(k)$.

④ Both *algorithm* 1 based on *fact* 4 and *algorithm* 2 based on *table* 1 are disordered & wrong logically.

⑤ To achieve so-called "breaking", the *example* in [8] was woven elaborately, and *table* 2 was falsified, namely its authors intendedly mutilated the two tuple data to cause indeterminacy.

⑥ The inverse $T^{-1}$ % $\bar{M}$ does not exist, and $Q^{-1}$ % $\bar{M}$ not necessarily exist.

Additionally, the case of $\Omega = \{5 + \delta, \ldots, (n+4) + \delta \mid \delta \geq n - 4\}$ with $f(i) + f(j) \neq f(k) \; \forall \; i, j, k \in [1, n]$ is not analyzed at all.

Therefore, the cryptanalysis of the REESSE1+ cryptosystem by [8] is a type of pseudo-attack and balderdash leading to which the most radical reason is that the authors of [8] are not aware of the indeterminacy of the lever function $\ell(.)$ namely $f(.)$, as is mentioned in [4]:

① if the order of $W$ is $d < \bar{M}$, then there is $W^{f(x)} \equiv W^{f(x)+d}$ (% $M$), and when $f(i) + f(j) = f(k)$, we see that $f(i) + d + f(j) + d \neq f(k) + d$;

② when $f(i) + f(j) \neq f(k)$, there exist $C_i \equiv A'_i W'^{f'(i)}$, $C_j \equiv A'_j W'^{f'(j)}$, and $C_k \equiv A'_k W'^{f'(k)}$ (% $M$) such that $f'(i) + f'(j) \equiv f'(k)$ (% $\bar{M}$) with $A'_k \leq \acute{p}$, where $\acute{p}$ is the maximal prime allowed.

Another vital reason is that [8] always regarded necessary conditions for $f(i) + f(j) = f(k)$ as sufficient and necessary conditions, and [8] did not consider the whole space of private keys or public keys.

## 2  Theorem 1 vs the REESSE1+ Cryptosystem

### 2.1  Condition at Theorem 1 in [8] Dissatisfies Necessity

Theorem 1 in [8] is retailed as follows:

**Theorem 1**: Let $\alpha$ be a real number, and let $r/s$ be a rational with $\gcd(r,s)=1$ and $|\alpha - r/s| < 1/(2s^2)$. Then $r/s$ is a convergent of the continued fraction expansion of $\alpha$.    ※

Here, $|\alpha - r/s|$ represents the absolute value of $(\alpha - r/s)$.

The proof of theorem 1 is referred to [9].

The condition $|\alpha - r/s| < 1/(2s^2)$ is only sufficient for $r/s$ to be a convergent of the continued fraction of $\alpha$, but not necessary. Namely if $r/s$ is a convergent of the continued fraction of $\alpha$, $|\alpha - r/s| < 1/(2s^2)$ does not necessarily hold.

An example is taken.

Example 1.

Let $r/s = 2/13$, and then $1/2s^2 = 1/(2 \times 13^2) = 0.002958579882$.

Let $\alpha = 2039/13001$, and then

$$2039/13001 - 2/13 = 0.002987935839 > 0.002958579882 = 1/(2 \times 13^2).$$

On the other hand, the continued fraction of $2039/13001$ is $1/(6 + (1/(2 + 1/(1 + \ldots 1/3))))$.

Thus, $2/13$ is a convergent of the continued fraction of $2039/13001$, which illustrates $|\alpha - r/s| < 1/(2s^2)$ is not necessary for $r/s$ to be a convergent of the continued fraction of $\alpha$.

## 2.2  $A_k$ Will Emerge But Is Undecidable If $f(i) + f(j) = f(k)$

Assume that $\acute{p}$ is the maximum prime in the cryptosystem, $\{A_1, \ldots, A_n\}$ is a coprime sequence with $0 < \forall A_x \leq \acute{p}$, $M > \prod_{x=1}^{n} A_x$ is a prime, and $C_x \equiv A_x W^{f(x)} \ (\% \ M)$ for $x = 1, \ldots, n$ is a public key [4], where $n \geq 6$, and $f(x) \in \Omega = \{5\delta, \ldots, (n+4)\delta \mid \delta = 1\} = \{5, \ldots, n+4\}$.

Assume $f(k) = f(i) + f(j)$ with $i \neq k$, $j \neq k$, and $i, j, k \in [1, \ldots, n]$. Let

$$Z \equiv C_i C_j C_k^{-1} \ (\% \ M).$$

Then

$$Z \equiv A_i A_j (A_k)^{-1} \ (\% \ M)$$

$$Z(A_k) \equiv A_i A_j \ (\% \ M)$$

$$Z(A_k) - LM = A_i A_j,$$

where $L$ is a positive integer.

Dividing the either side of the above equation by $(MA_k)$ yields

$$Z/M - L/A_k = A_i A_j / (M A_k). \qquad (1)$$

Due to $M > \prod_{x=1}^{n} A_x$ and every $A_x \geq 2$, we have

$$Z/M - L/A_k < 1/(2^{n-2-1} A_k^2). \qquad (1')$$

Obviously, when $n > 2 + 1$, (1′) may have a variant, namely

$$Z/M - L/A_k < 1/(2 A_k^2). \qquad (1'')$$

In terms of theorem 1, $L/A_k$ is a convergent of the continued fraction of $Z/M$.

Let $p_0/q_0, p_1/q_1, \ldots, p_t/q_t$ be the convergent sequence of continued fraction of $Z/M$, and

$$L/A_k = p_u/q_u.$$

Note that if $p_u/q_u$ satisfies (1″), then $p_{u+1}/q_{u+1}$, $p_{u+2}/q_{u+2}$, …, $p_t/q_t$ also likely satisfies (1″). Therefore, there likely exist multiple values of $L/A_k$ by (1″), and $A_k$ is undetermined.

However, if we do not know in advance whether $f(i) + f(j) = f(k)$, then even if $Z/M - p_u/q_u < 1/(2q_u^2)$, we can not decide $f(i) + f(j) = f(k)$. Namely $Z/M - p_u/q_u < 1/(2q_u^2)$ is only necessary for $f(i) + f(j) = f(k)$, but not sufficient, which will be discussed further in what follows.

## 3  Conditions at *Fact* 1.1 and 4 Each Are Insufficient for $f(i) + f(j) = f(k)$

Because *fact* 4 is essentially equivalent to each of *fact* 1.2 and 1.3, if the condition at *fact* 4 is insufficient for $f(i)+f(j) = f(k)$, the conditions at *fact* 1.2 and 1.3 each are also insufficient.

The condition $q_{u+1} > q_u \Delta = q_u(M / (2\prod_{x=n-2}^{m} prime\langle x\rangle))^{1/2}$ at *fact* 4 connotes (1″) at *fact* 1.1 because (1″) is the precondition of $q_{u+1} > q_u \Delta$ which is the dominant basis of *alg*.1 [8].

## 3.1  Converse Proposition of *Fact* 1.1 does not Hold and (1″) Is Insufficient for $f(i) + f(j) = f(k)$

*Fact* 1.1 in [8] is retailed as follows:

**Fact** 1.1 [8]: If $f(i) + f(j) = f(k)$, there exists a $q_u$ such that $q_u = A_k$ in $\{p_0 / q_0, p_1 / q_1, \ldots, p_t / q_t\}$, the convergent sequence of continued fraction expansion of $Z / M$ with $Z \equiv C_i C_j C_k^{-1} \% M$.    ※

Due to $f(i) + f(j) = f(k)$, $Z / M = L / A_k + A_i A_j / (M A_k)$, $M > \prod_{x=1}^{n} A_x$ and $A_x \geq 2$, we have
$$Z / M - L / A_k = A_i A_j / (M A_k) < A_i A_j / (A_k \prod_{x=1}^{n} A_x).$$

Further,
$$Z / M - L / A_k < 1 / (2 A_k^2). \qquad (1'')$$

Let $Z / M = [0; a_1, a_2, \ldots, a_t]$ is the continued fraction expansion of $Z/M$.

By theorem 1, $\exists u \in [1, t]$ makes $Z/M - p_u/q_u < 1/(2 q_u^2)$.

Let $L / A_k = p_u / q_u$, where
$$p_u / q_u = a_0 + 1 / (a_1 + 1 / (a_2 + \ldots + 1 / (a_{u-1} + 1 / a_u))). \qquad (2)$$

Notice that it is possible that $\exists h > 0$ makes $Z/M - p_{u+h}/q_{u+h} < 1/(2 q_{u+h}^2)$, and moreover not *fact* 1.1 but its converse is the inner logical base of *alg*.1 in [8].

Through a counterexample, we will prove that the converse proposition of *fact* 1.1 does not hold, that is, the condition $Z/M - p_u/q_u < 1/(2 q_u^2)$ is insufficient for $f(i) + f(j) = f(k)$.

Example 2.

For convenience in computing, let $n = 6$, $\{A_x\} = \{11, 10, 3, 7, 17, 13\}$, $\delta = 1$, and $M = 510931$.

Arbitrarily select $W = 17797$, $f(1) = 9$, $f(2) = 6$, $f(3) = 10$, $f(4) = 5$, $f(5) = 7$, and $f(6) = 8$.

From $C_x \equiv A_x W^{f(x)} \ (\% M)$, we obtain

$\{C_x\} = \{113101, 79182, 175066, 433093, 501150, 389033\}$,

and its inverse sequence

$\{C_x^{-1}\} = \{266775, 236469, 435654, 149312, 434038, 425203\}$.

Randomly select $i = 1$, $j = 3$, and $k = 5$. In this case, $f(5) = 7 \neq f(1) + f(3) = 9 + 10$. Compute
$$Z \equiv C_1 C_3 C_5^{-1}$$
$$\equiv 113101 \times 175066 \times 434038$$
$$\equiv 186640 \ (\% \ 510931).$$

Presume that $W$ in $C_1 C_3$ is just neutralized by $W^{-1}$ in $C_5^{-1}$, then
$$186640 \equiv A_1 A_3 A_5^{-1} \ (\% \ 510931).$$

According to (1),
$$186640 / 510931 - L / A_5 = A_1 A_3 / (510931 A_5).$$

By the Euclidean algorithm, $a_1, a_2, a_3, \ldots$ are found out, and thus the continued fraction of
$$186640 / 510931 = 1/(2 + 1 / (1 + 1 / (2 + 1 / (1 + 1 / (4 + \ldots + 1 / 3))))).$$

Heuristically let
$$p_4 / q_4 = L / A_5 = 1 / (2 + 1 / (1 + 1 / (2 + 1 / 1))) = 4 / 11,$$

which indicates that probably $A_5 = 11$. On this occasion, there is

$$186640 / 510931 - 4 / 11 = 0.0016575801$$
$$< 1 / (2 A_5^2) = 1 / (2 \times 11^2) = 0.0041322314.$$

The above expression satisfies (1″), namely the condition at theorem 1, and thereby $A_5 = 11$ less than the maximum in $\{A_x\}$ is deduced, which is in direct contradiction to factual $A_5 = 17$.

So the condition $Z / M - p_u / q_u < 1 / (2 q_u^2)$ is not sufficient for $f(i) + f(j) = f(k)$, namely the converse proposition of *fact* 1.1 does not hold.

### 3.2   Each of *Fact* 1.2, 1.3 and 4 does not Hold

*Fact* 1.2 in [8] is retailed as follows:

**Fact 1.2** [8]: There is sharp increase from $q_u$ to $q_{u+1}$ since $q_{u+1} \geq (A_k M / (2A_i A_j))^{1/2}$.

The derivation of *fact* 1.2 in [8] is retailed as follows:

Let $L / A_k$ be the $u$-th convergent, i.e., $q_u = A_k$ and $p_u = L$, i.e., $p_u / q_u = L / A_k$. Then we know that

$$|Z / M - p_{u+1} / q_{u+1}| < A_i A_j / (A_k M) = 1 / (2((A_k M / (2A_i A_j))^{1/2})^2). \tag{2′}$$

According to theorem 1 and convergence of sequence $\{p_0 / q_0, p_1 / q_1, \ldots, p_t / q_t\}$, we obtain that

$$q_{u+1} \geq (A_k M / (2 A_i A_j))^{1/2} = A_k (M / (2 A_i A_j A_k))^{1/2}. \tag{3}$$

※

Is the above derivation right? See the following analysis.

Clearly, by the definition of a finite continued fraction, (2′) holds. In addition, in terms of [9], $p_{u+1}$ are $q_{u+1}$ are coprime, and there is $q_{u+1} \geq A_k = q_u$, which is a judgment foundation.

If $f(i) + f(j) = f(k)$, then there is $|Z / M - p_u / q_u| < 1/(2 q_u^2)$ with $L / A_k = p_u / q_u$. Furthermore, through practical observations, in most cases, there is also

$$|Z / M - p_{u+1} / q_{u+1}| < 1 / (2 q_{u+1}^2). \tag{3′}$$

According to (2′) and (3′), we have either

$$|Z / M - p_{u+1} / q_{u+1}| < 1 / (2 q_{u+1}^2) < 1 / (2 ((A_k M / (2 A_i A_j))^{1/2})^2), \tag{3″}$$

or

$$|Z / M - p_{u+1} / q_{u+1}| < 1 / (2 ((A_k M / (2 A_i A_j))^{1/2})^2) < 1 / (2 q_{u+1}^2). \tag{3‴}$$

If (3″) holds, there exists $q_{u+1} \geq A_k (M / (2 A_i A_j A_k))^{1/2}$, which also indicates $q_{u+1} \geq A_k = q_u$.

If (3‴) holds, there exists $A_k (M / (2 A_i A_j A_k))^{1/2} \geq q_{u+1}$. Notice that in this case, $q_{u+1} \geq A_k = q_u$ is still possible.

Therefore, $q_{u+1} \geq A_k (M / (2 A_i A_j A_k))^{1/2}$, namely *fact* 1.2 does not necessarily hold, which indicates that there is a logic error during the derivation of (3) in [8].

Moreover, from (2′) and (3′) we can judge that when $n$ is large enough — 80 for example, the probability that (3‴) holds is greater than one that (3″) holds.

Now, we review *fact* 1.3 in [8]. It is retailed as follows:

**Fact 1.3** [8]: Due to *fact* 1.2, there is also a sharp increase from $a_u$ to $a_{u+1}$, since $q_{v+1} = a_{v+1} q_v + q_{v-1}$ for $v = 1, 3, \ldots, t$. Here $a_v$s are items of $Z / M$ determined by (2).       ※

Obviously, because *fact* 1.2 does not hold, *fact* 1.3 does not also hold.

Further, because *fact* 1.2, namely (3) does not hold, naturally, *fact* 4 in [8] does not also hold, that is, $q_{u+1}$

$> q_u (M / (2\prod_{x=n-2}^{m} prime\langle x\rangle))^{1/2}$ is not always valid.

Observe an example once more.

In example 3, suppose that the bit-length of a plaintext block is 8, and two bits of a block correspond to three items of a coprime sequence $\{A_x\}$, which means that the encryption algorithm is optimized through a compact binary sequence. In practice, we do just so.

Apparently, the length of $\{A_x\}$ is $3 \times (8 / 2) = 12$.

Example 3.

Let $\{A_x\}$ = {{23, 11, 17}, {41, 29, 26}, {15, 19, 37}, {31, 7, 43}}, and

$M = 2022169 > 31 \times 37 \times 41 \times 43 = 2022161$.

Randomly select $W = 1507351$, $f(1) = 6$, $f(2) = 14$, $f(3) = 9$, $f(4) = 11$, $f(5) = 12$, $f(6) = 10$, $f(7) = 8$, $f(8) = 16$, $f(9) = 5$, $f(10) = 13$, $f(11) = 15$, and $f(12) = 7$.

From $C_x \equiv A_x W^{f(x)}$ (% $M$), we obtain $\{C_x\}$ =

{{572402, 1930240, 374715}, {25128, 265158, 350520},
{1674837, 1231458, 1448214}, {110225, 1198155, 757620}},

and $\{C_6^{-1}, C_7^{-1}\}$ = {93176, 1591882}. Let

$$Z \equiv (C_4 C_{12})(C_6^{-1} C_7^{-1})$$
$$\equiv (25128 \times 757620)(93176 \times 1591882)$$
$$\equiv 776394 \times 1123251$$
$$\equiv 689616 \ (\% \ 2022169).$$

Then, $689616 / 2022169 - L / (A_6 A_7) = (A_4 A_{12}) / (2022169 A_6 A_7)$.

Further, the continued fraction of $689616 / 2022169$ is

$1 / (2 + 1 / (1 + 1/ (13 + 1/ (1 + (1/ (3 + 1/ (2 + 1/ (2 + 1/ (2 + 1/ (97 + 4 / 9))))))))))$.

Heuristically let

$$L / (A_6 A_7) = 1 / (2 + 1 / (1 + 1/ (13 + 1/ (1 + (1/ (3 + 1 / 2))))))$$
$$= 133 / 390,$$

which indicates that probably $A_6 A_7 = 390$. Because the discriminant

$$689616 / 2022169 - 133 / 390 = 2.235477262e-6$$
$$< 1 / (2 \times 390^2) = 3.287310979e-6$$

satisfies the condition at theorem 1 in [8], $A_6 A_7 = 390$ is deduced out.

The integer 390 may be factorized into the pairs (2, 195), (3, 130), (5, 78), (6, 65), (10, 39), (13, 30), or (15, 26), where the elements of (10, 39), (13, 30), and (15, 26) are less than maximal number in $\{A_x\}$. Thus, true $(A_6, A_7) = (26, 15)$ is included in 6 potential cases. Here, $a_u = 2$ and also $a_{u+1} = 2$, and there is no sharp increase from $a_u$ to $a_{u+1}$.

Additionally, this example also illustrates that when one attempts to infer the suitable factors of the product $A_{k_1} A_{k_2}$ by $f(i) + f(j) = f(k_1) + f(k_2)$ with every $f(x) \in \Omega = \{n + 1, \ldots, 2n\}$, indeterminacy is increased remarkably.

### 3.3  Condition at *Fact* 4 Is Insufficient for $f(i) + f(j) = f(k)$

In [6], the attackers attempted to seek $A_k$ dominantly by the converse proposition of *fact* 1.1, and however, disturbing values of $A_k$ are too many to determine the original value of $A_k$. Therefore, in [8], the attackers attempted to diminish indeterminacy of $A_k$ through *fact* 4 which connotes *fact* 1.1, and is equivalent to each of *fact* 1.2 and 1.3.

To say the least, even if *fact* 4 is valid sometimes, we can prove by a counterexample that the condition at *fact* 4 is insufficient for $f(i) + f(j) = f(k)$.

Example 4.

Still let $n = 6$, $\{A_x\} = \{11, 10, 3, 7, 17, 13\}$, and $M = 510931 > 11 \times 10 \times 3 \times 7 \times 17 \times 13 = 510510$.

Arbitrarily select $W = 17797$, $f(1) = 9$, $f(2) = 6$, $f(3) = 10$, $f(4) = 5$, $f(5) = 7$, and $f(6) = 8$.

From $C_x \equiv A_x W^{f(x)}$ (% $M$), we obtain $\{C_x\} = \{113101, 79182, 175066, 433093, 501150, 389033\}$, and its inverse sequence $\{C_x^{-1}\} = \{266775, 236469, 435654, 149312, 434038, 425203\}$.

Randomly select $i = 1, j = 3$, and $k = 6$. In this case, $f(6) = 8 \neq f(1) + f(3) = 9 + 10$. Compute

$$Z \equiv C_1 C_3 C_6^{-1}$$
$$\equiv 113101 \times 175066 \times 425203$$
$$\equiv 425865 \ (\% \ 510931).$$

Presume that $W$ in $C_1 C_3$ is just neutralized by $W^{-1}$ in $C_6^{-1}$, then

$$425865 \equiv A_1 A_3 A_6^{-1} \ (\% \ 510931).$$

According to *alg*.1 in [8],

$$425865 \ / \ 510931 - L \ / \ A_6 = A_1 A_3 \ / \ (510931 \ A_6).$$

Compute the continued fraction of $186640 / 510931$ being

$$1 / (1 + 1 / (5 + 1 / (159 + 1 / 535))).$$

Heuristically let

$$L / A_6 = 1 / (1 + 1 / 5) = 5 / 6,$$

which indicates that probably $A_6 = 6$. Further, can verify that

$$425865 / 510931 - 5 / 6 = 0.000174518$$
$$< 1 / (2 \times 6^2) = 0.0138889$$

satisfies the condition at theorem 1 in [8].

Let $u = 2$, and $q_u = A_k = A_6 = 6$.

Then $p_{u+1} / q_{u+1} = p_3 / q_3 = 1 / (1 + 1 / (5 + 1 / 159)) = 796 / 955$, and

$$A_k (M / (2 A_i A_j A_k))^{1/2} = 6 \ (510931 / (2 \times 11 \times 3 \times 6))^{1/2}$$
$$= 6 \times 35.9197$$
$$= 215.5186.$$

In addition, evidently $prime\langle 1\rangle = 2$, $prime\langle 2\rangle = 3$, $prime\langle 3\rangle = 5$, $prime\langle 4\rangle = 7$, $prime\langle 5\rangle = 11$, $prime\langle 6\rangle = 13$, $prime\langle 7\rangle = 17$, and $prime\langle 8\rangle = 19$ which are according to [8].

Then, by *fact* 4 in [8], $m = 7$, and $\Delta = (M / (2 \prod_{x=n-2}^{m} prime\langle x\rangle))^{1/2} = (15)^{1/2} = 3.8729$.

Thus, $q_{u+1} = 955 > A_k (M / (2 A_i A_j A_k))^{1/2} = 216$ satisfies *fact* 1.2 namely (3), $a_{u+1} = 159 > a_u = 5$ satisfies *fact* 1.3, and $q_{u+1} = 955 > q_u \Delta \approx 24$ satisfies *fact* 4 and *alg*.1.

By the condition at *fact* 4, $A_6 = 6 < \max A = 221$ is deduced, namely *alg*.1 will output $\{1, 3, 6, 6\}$. However, it is in direct contradiction to true $A_6 = 13$, which show the condition at *fact* 4 is not sufficient for $f(i) + f(j) = f(k)$, and every $A_x$ will likely be evaluated to at least two eligible values (see example 5).

Because the condition at *fact* 4 is insufficient for $f(i) + f(j) = f(k)$, property 1, 2, 3, 4 and 5 are invalid. Further, the run result of *alg*.1 regarding an arbitrary public key $\{C_1, …, C_n\}$ as an input will contain enormous disturbing data as $n \geq 80$, and it is infeasible that *alg*.2 find out the original coprime sequence $\{A_x\}$ in polynomial time (see example 5), which manifests that *alg*.1 and 2 are invalid.

## 4  Example in [8] Is Woven Elaborately and Data at *Table* 2 Is Falsified

### 4.1  Example in [8] Illustrates Nothing about Breaking

It is easily understood that according to *fact* 1.1 and 4, the authors of [8] can weave an example consistent with *alg*.1 and 2 since

① the lever function value $\{f(1), …, f(n)\}$ may be known in advance;

② the coprime sequence $\{A_1, …, A_n\}$ may be selected elaborately in advance;

③ the condition $Z / M - L / A_k < 1 / (2 A_k^2)$ at *fact* 1.1 is necessary for $f(i) + f(j) = f(k)$;

④ the condition $q_{u+1} > q_u (M/(2\prod_{x=n-2}^{m} prime \langle x \rangle))^{1/2}$ at *fact* 4 is necessary for $f(i)+f(j)=f(k)$ sometime.

However, as is indicated in the above rebuttal, a consistent example does not illustrates that a related $\{A_x\}$ can be extracted accurately from an arbitrary public key $\{C_x\}$ when $\{f(x)\}$ and $\{A_x\}$ are unknown in advance. The authors of [8] at most broke "their own REESSE1+", which diverted themselves, but not our REESSE1+ with choice parameters. It is well understood that even though a cryptosystem is RSA or ECC, its parameter is must also selected; otherwise the cryptosystem is insecure.

The example in [8] is neither readable nor verifiable in short time, and the proportion of $n$ to $\log_2 M$ is not also proper, which contravenes the optimization principle for the modulus $M$ in the REESSE1+ cryptosystem. An obvious truth is that if $M$ is too large, the length of a public key will increase rapidly. Therefore, $M$ should be as small as possible while at least meets $M > \prod_{x=1}^{n} A_x$ meantime. Selection of the sequence $\{A_x\}$ in [8] also contravenes the optimization principle.

The intent for [8] to select such a large $M$ that $n$ is out of proportion to $\log_2 M$ seems to want to increase the necessity of the conditions at *fact* 1.1 and 4 for $f(i) + f(j) = f(k)$. However, it can not increase the sufficiency of the conditions.

### 4.2  Data at *Table* 2 Is Falsified for a Compatible Effect

In above paragraphs, we illustrate that the condition $Z / M - L / A_k < 1 / (2 A_k^2)$ at *fact* 1.1, namely (1″) is insufficient for $f(i) + f(j) = f(k)$. Property I will make us better understand it.

**Property I**: Let $C_x \equiv A_x W^{f(x)}$ (% $M$), where every $x \in [1, n]$, $A_x \leq \acute{p}$, $f(x) \in \{5, …, n + 4\}$, $M > \prod_{x=1}^{n} A_x$ is a prime. Then, $\forall i, j, k \in [1, n]$, even if $f(i)+f(j) \neq f(k)$,

1) there always exist

$$C_i \equiv A'_i W'^{f'(i)}, C_j \equiv A'_j W'^{f'(j)}, \text{ and } C_k \equiv A'_k W'^{f'(k)} \text{ (% } M\text{)}$$

such that $f'(i) + f'(j) \equiv f'(k)$ (% $\overline{M}$) with $A'_k \leq \acute{p}$.

2) $C_i, C_j, C_k$ make (1″) hold with $A'_k \leq \acute{p}$ in all probability.

*Proof:*

1)

Let $O_d$ be an oracle for a discrete logarithm.

Suppose that $W' \in [1, \overline{M}]$ is a generator of $(\mathbb{Z}_M^*, \cdot)$.

In terms of group theories, $\forall\, A'_k \in \{2, \ldots, \acute{p}\}$, the equation

$$C_k \equiv A'_k W'^{f'(k)} \ (\%\, M)$$

has a solution. $f'(k)$ may be taken through $O_d$.

$\forall\, f'(i) \in [1, \overline{M}]$, and let $f'(j) \equiv f'(k) - f'(i) \ (\%\, \overline{M})$.

Then, from $C_i \equiv A'_i W'^{f'(i)}$ and $C_j \equiv A'_j W'^{f'(j)} \ (\%\, M)$, we can obtain many distinct pairs $(A'_i, A'_j)$, where $A'_i, A'_j \in (1, M)$, and $f'(i) + f'(j) \equiv f'(k) \ (\%\, \overline{M})$.

2)

Let

$$Z \equiv C_i C_j C_k^{-1} \equiv A'_i A'_j W'^{f'(i)+f'(j)} (A'_k W'^{f'(k)})^{-1} \ (\%\, M)$$

with $f'(i) + f'(j) \equiv f'(k) \ (\%\, \overline{M})$ but $f(i) + f(j) \neq f(k)$.

Further, there is $A'_i A'_j \equiv C_i C_j C_k^{-1} A'_k \ (\%\, M)$.

It is easily seen from the above equations the values of $W'$ and $f'(k)$ do not influence the value of $A'_i A'_j$.

If $A'_k \in [2, \acute{p}]$ changes, $A'_i A'_j$ also changes. Thus, $\forall\, i, j, k \in [1, n]$, the number of value of $A'_i A'_j$ is $\acute{p} - 1$.

Let $M = 2q\acute{p}^2 A'_k$, where $q$ is a rational number.

According to (1),

$$Z/M - L/A'_k = A'_i A'_j / (M A'_k) = A'_i A'_j / (2q\acute{p}^2 A'^2_k).$$

When $A'_i A'_j \leq q\acute{p}^2$, there is

$$Z/M - L/A'_k \leq q\acute{p}^2 / (2q\acute{p}^2 A'^2_k) = 1/(2 A'^2_k)$$

which satisfies (1″).

Assume that the value of $A'_i A'_j$ distributes uniformly on $(1, M)$. Then, the probability that $A'_i A'_j$ makes (1″) hold is

$$\begin{aligned}P_{\forall\, i,\, j,\, k\, \in [1,\, n]} &= (q\acute{p}^2 / (2q\acute{p}^2))\,(1/2 + \ldots + 1/\acute{p}) \\ &\geq (1/2)(2(\acute{p}-1)/(\acute{p}+2)) \\ &= 1 - 3/(\acute{p}+2).\end{aligned}$$

It is seen that the probability is very large.    □

According to property I.2, for a certain $C_k \in \{C_1, \ldots, C_n\}$ and $\forall\, C_i, C_j \in \{C_1, \ldots, C_n\}$, $A_k$ will have roughly $n^2$ values by (1″) namely the condition at *fact* 1.1, including the repeated, and considering the symmetry, almost every value has at least one counterpart.

Of course, if the condition at *fact* 4, namely $q_{u+1} > q_u \Delta$ which connotes (1″) is used as a constraint, the number of values of $A_k = q_u$ will decrease. Example 4 already shows that even though $f(i) + f(j) \neq f(k)$, an eligible $A_k$ can still be found.

Notice that when $i, j, k$ all fix on, it is fully possible that $L/A_k$ has multiple satisfactory values, which implies multiple convergents of the continued fraction of $Z/M$ likely meet (1″) and even $q_{u+1} > q_u \Delta$.

To clarify the matter thoroughly, we program by *alg*.1 in MS Visual C++, make an executable file, repeat the experiment regarding the public key at the example in [8] as input, and obtain the following output which is classified the same as in [8]:

| $A_k$ | Tuples $(i, j, k)$ |
|---|---|
| $A_1 = 9$ | (9, 9, 1) |
| $A_2 = 253$ | (7, 5, 2), (9, 6, 2), (5, 7, 2), (6, 9, 2) |
| $A_3 = 16127$ | (10, 7, 3), (7, 10, 3) |
| $A_4 = 3$ | (8, 3, 4), (3, 8, 4) |
| $A_4 = 205$ | (9, 3, 4), (6, 5, 4), (5, 6, 4), (7, 7, 4), (3, 9, 4) |
| $A_4 = 152391460756$ | (8, 7, 4), (7, 8, 4) |
| $A_6 = 53022327$ | (4, 3, 6), (3, 4, 6) |
| $A_6 = 318461273008612$ | (4, 3, 6), (3, 4, 6) |
| $A_6 = 4471789987666990$ | (5, 3, 6), (3, 5, 6) |
| $A_6 = 1572955621791218$ | (5, 5, 6) |
| $A_8 = 2809$ | (5, 5, 8), (9, 7, 8), (7, 9, 8) |
| $A_{10} = 49$ | (9, 5, 10), (5, 9, 10) |
| $A_{10} = 1894$ | (9, 6, 10), (6, 9, 10) |
| $A_{10} = 6957$ | (9, 7, 10), (7, 9, 10) |

Table I: Output of the program by *alg*. 1 given the public key at the example in [8]

Obviously, *table* 2 in [8] misrepresented $A_3 = 16127$ as $A_4 = 16127$, and $A_6 = 53022327$ as $A_{10} = 53022327$. What gets worse is that *table* 2 mutilated the two tuple data (4, 3, 6, 318461273008612) and (3, 4, 6, 318461273008612), which is a type of data falsification. These two tuple data illustrate that for fixed *i*, *j*, *k*, the $L / A_k$ may have several satisfactory values, namely the several convergents of the continued fraction of $Z / M$ meet *fact* 4 meantime, which reflects the insufficiency of the condition $q_{u+1} > q_u \Delta$ further, increases the indeterminacy of $A_k$ greatly, and weakens the reliability of *alg*.1 in [8] greatly.

### 4.3  Example in [8] Is Woven Elaborately and Alg.2 in [8] Is Invalid

In the above, it is mentioned that at most the authors of [8] broke "their own REESSE1+", because the example in [8] is woven elaborately, and the parameters $\{A_x\}$ and $\{f(x)\}$ are selected deliberately.

If we use another set of parameters for producing a public key as the input of the program by *alg*.1, the output result will contains so many disturbing data that the original sequence $\{A_1, …, A_n\}$ can not be distinguished in polynomial time.

Example 5.

Let $n = 10$, $\{A_x\} = \{437, 221, 77, 43, 37, 29, 41, 31, 15, 2\}$, and

$M = 13082761331670077 > \prod_{x=1}^{n} A_x = 13082761331670030$.

Arbitrarily select $W = 944516391$, $f(1) = 11$, $f(2) = 14$, $f(3) = 13$, $f(4) = 8$, $f(5) = 10$, $f(6) = 5$, $f(7) = 9$, $f(8) = 7$, $f(9) = 12$, $f(10) = 6$.

According to $C_x \equiv A_x W^{f(x)} \ (\% \ M)$, we obtain $\{C_x\} = \{3534250731208421, 12235924019299910, 8726060645493642, 10110020851673707, 2328792308267710, 8425476748983036, 6187583147203887, 10200412235916586, 9359330740489342, 5977236088006743\}$.

Input the public key $\{C_x\}$ into the program by *alg*.1, and obtain $\Delta = 506$, max$A = 58642670$, and the following tuples greater than 100:

| $A_k$ | Tuples $(i, j, k)$ |
|---|---|
| $A_1 = 187125$ | (1, 1, 1) |
| $A_1 = 121089$ | (2, 1, 1), (1, 2, 1) |
| $A_1 = 77$ | (5, 3, 1), (3, 5, 1) |
| $A_1 = 23$ | (8, 6, 1), (6, 8, 1), (10, 10, 1) |
| $A_1 = 437$ | (10, 6, 1), (6, 10, 1) |
| $A_2 = 1251$ | (1, 1, 2) |
| $A_2 = 187125$ | (2, 1, 2), (1, 2, 2) |
| $A_2 = 121089$ | (2, 2, 2) |
| $A_2 = 17$ | (8, 4, 2), (6, 5, 2), (5, 6, 2), (10, 7, 2), (4, 8, 2), (7, 10, 2) |
| $A_2 = 221$ | (10, 4, 2), (7, 6, 2), (6, 7, 2), (8, 8, 2), (4, 10, 2) |
| $A_2 = 77$ | (9, 8, 2), (8, 9, 2) |
| $A_2 = 4204$ | (10, 10, 2) |
| $A_3 = 187125$ | (3, 1, 3), (1, 3, 3) |
| $A_3 = 12$ | (7, 1, 3), (1, 7, 3) |
| $A_3 = 121089$ | (3, 2, 3), (2, 3, 3) |
| $A_3 = 77$ | (6, 4, 3), (4, 6, 3), (10, 8, 3), (8, 10, 3) |
| $A_3 = 11$ | (10, 4, 3), (7, 6, 3), (6, 7, 3), (8, 8, 3), (4, 10, 3) |
| $A_3 = 2113$ | (8, 7, 3), (7, 8, 3) |
| $A_3 = 769$ | (9, 8, 3), (8, 9, 3) |
| $A_4 = 187125$ | (4, 1, 4), (1, 4, 4) |
| $A_4 = 121089$ | (4, 2, 4), (2, 4, 4) |
| $A_4 = 76$ | (10, 6, 4), (6, 10, 4) |
| $A_4 = 56$ | (10, 9, 4), (9, 10, 4) |
| $A_5 = 187125$ | (5, 1, 5), (1, 5, 5) |
| $A_5 = 630269$ | (6, 1, 5), (1, 6, 5) |
| $A_5 = 121089$ | (5, 2, 5), (2, 5, 5) |
| $A_5 = 41$ | (8, 2, 5), (2, 8, 5) |
| $A_5 = 97$ | (4, 3, 5), (3, 4, 5) |
| $A_5 = 37$ | (6, 6, 5), (10, 6, 5), (6, 10, 5) |
| $A_6 = 187125$ | (6, 1, 6), (1, 6, 6) |
| $A_6 = 121089$ | (6, 2, 6), (2, 6, 6) |
| $A_7 = 187125$ | (7, 1, 7), (1, 7, 7) |
| $A_7 = 121089$ | (7, 2, 7), (2, 7, 7) |
| $A_7 = 3$ | (9, 3, 7), (3, 9, 7) |
| $A_8$ = 187125 | (8, 1, 8), (1, 8, 8) |
| $A_8$ = 34945619 | (6, 2, 8), (2, 6, 8) |
| $A_8$ = 121089 | (8, 2, 8), (2, 8, 8) |
| $A_9$ = 187125 | (9, 1, 9), (1, 9, 9) |
| $A_9$ = 121089 | (9, 2, 9), (2, 9, 9) |
| $A_9$ = 5 | (6, 4, 9), (4, 6, 9), (10, 8, 9), (8, 10, 9) |
| $A_9$ = 15 | (8, 6, 9), (6, 8, 9), (10, 10, 9) |
| $A_{10}$ = 259970 | (4, 1, 10), (1, 4, 10) |
| $A_{10}$ = 187125 | (10, 1, 10), (1, 10, 10) |
| $A_{10}$ = 121089 | (10, 2, 10), (2, 10, 10) |
| $A_{10}$ = 7629 | (8, 3, 10), (3, 8, 10) |

Table II: Output of the program by *alg*. 1 given the public key at example 5

From table II, we observe that

$A_k$ from 5 tuples is $A_2$ = 221 or $A_3$ = 11,

$A_k$ from 4 tuples is $A_3$ = 77 or $A_9$ = 5,

$A_k$ from 3 tuples is $A_1$ = 23, $A_5$ = 37, or $A_9$ = 15,

$A_k$ from 2 tuples is $A_1$ = 77, $A_2$ = 77, $A_3$ = 12, $A_4$ = 56, $A_5$ = 41, or $A_7$ = 3 etc, and

$A_k$ from 1 tuples is $A_1$ = 187125, $A_2$ = 1251, $A_2$ = 121089, or $A_2$ = 4204.

Among these $A_k$'s, there exist at least $2^{n-5}$ compatible combinations.

For instance, arbitrarily select compatible $A_3$ = 11, $A_9$ = 5, $A_1$ = 23, $A_5$ = 41, and $A_2$ = 1251, and find out $f(3) = 14, f(9) = 13, f(1) = 12, f(5) = 11$, and $f(2) = 10$ by Table 1 in [8].

Again for instance, arbitrarily select compatible $A_3$ = 11, $A_9$ = 5, $A_5$ = 37, $A_7$ = 3, and $A_1$ = 187125, and find out $f(3) = 14, f(9) = 13, f(5) = 12, f(7) = 11$, and $f(1) = 10$ by Table 1 in [8].

Therefore, if keep $\Omega = \{5, .., n + 4\}$ unvaried, we may select fit $\{A_x\}$ and $W$ so as to make the time complexity of the continued fraction attack by $q_{u+1} > q_u \Delta$ and *table* 1 get to at least O($2^n$), which elucidates that the example woven elaborately in [8] has no practical meaning, and *alg*.2 in [8] is invalid.

However, we had best select fit $\Omega$ while let $\{A_x\}$ and $W$ random so as to avoid attack by (1′) (see sect.5.1).

### 4.4  Distribution of Tuples Relating $A_k$ does not Follow *Table* 1 in [8]

In addition, from table II we also observe that $A_2$ = 17 involves 6 tuples, and $A_5$ = 37 involves 3 tuples (but in fact, 6 tuples is impossible, and $f(5) = 10$), which indicates that the distribution of tuples relating $A_k$ does not follow *table* 1 in [8]. Besides, considering $A_3$ = 11 from 5 tuples, $A_9$ = 5 from 4 tuples etc, we see that *table* 1 is insufficient for $f(i) + f(j) = f(k)$, that is, the converse proposition of fact 2.2 does not hold.

## 5  Why Is $C_x \equiv A_x W^{f(x)}$ (% $M$) Changed to $C_x \equiv (A_x W^{f(x)})^\delta$ (% $M$) in REESSE1+ v2.1

### 5.1  Lever Set $\Omega$ Needs to Be Complicated When $C_x \equiv A_x W^{f(x)}$ (% $M$)

In REESSE1, $C_x \equiv A_x W^{f(x)}$ (% $M$) with $f(x) \in \Omega = \{5, …, n + 4\}$.

In REESSE1+, $C_x \equiv A_x W^{f(x)}$ (% $M$) with $f(x) \in \Omega = \{5\delta, \ldots, (n+4)\delta | \delta \geq 1\}$, $\{5 + \delta, \ldots, (n+4) + \delta | \delta \geq n - 4\}$, $\{5, 7, \ldots, 19, 53, 55, \ldots\}$ etc.

If let $W' = W^\delta$ (% $M$), we see that $\{5\delta, \ldots, (n+4)\delta | \delta \geq 1\}$ is substantially the same as $\{5, \ldots, n+4\}$.

Although [8] by $Z/M - L/A_k < 1/(2 A_k^2)$ and $q_{u+1} > q_u \Delta$ can not break REESSE1+ with $C_x \equiv A_x W^{f(x)}$ (% $M$) and $\Omega = \{5\delta, \ldots, (n+4)\delta | \delta \geq 1\}$, attack by $Z/M - L/A_k < 1/(2^{n-2-1} A_k^2)$, namely (1') will filter out the most of disturbing data as $n$ is large, which makes REESSE1+ be faced with danger. Therefore, in REESSE1+ with $C_x \equiv A_x W^{f(x)}$ (% $M$), $\Omega$ needs to be complicated, namely had best select $\Omega = \{5, 7, \ldots, 19, 53, 55, \ldots\}$ which is an odd set of $2n$ elements such that ① $\forall e_1, e_2 \in \Omega, e_1 \neq e_2$, ② $\forall e_1, e_2, e_3 \in \Omega, e_1 + e_2 \neq e_3$, ③ $\forall e_1, e_2, e_3, e_4 \in \Omega, e_1 + e_2 + e_3 \neq e_4$.

## 5.2 Key Transform $C_x \equiv A_x W^{f(x)}$ (% $M$) Needs to Be Strengthened When Still $\Omega = \{5, \ldots, n + 4\}$

In REESSE1+ with $C_x \equiv A_x W^{f(x)}$ (% $M$) and $f(x) \in \Omega = \{5, 7, \ldots, 19, 53, 55, \ldots\}$, because the elements of $\Omega$ are relatively large, decryption speed will decrease greatly.

To keep $\Omega = \{5, \ldots, n + 4\}$ unvaried, the key transform should be strengthened, so in REESSE1+ v2.1, we let $C_x \equiv (A_x W^{f(x)})^\delta$ (% $M$). In this way, REESSE1+ v2.1 is not only secure but also swift.

## 6 Attack on the Signature Is an Eisegesis

### 6.1 $T^{-1}$ % $\bar{M}$ does not Exist and $Q^{-1}$ % $\bar{M}$ not Necessarily Exist

Section 4 of the original [8] deduces $U \equiv ((Q/H)^{1/S} \hat{G}(GW)^{-1} \delta^{\delta(\delta+1) - 1/S})^{QT}$ (% $M$), which is right.

However, $(GW)^{-1} \delta^{\delta(\delta+1) - 1/S} \equiv ((Q/H)^{-S-1} \hat{G}^{-1}) U^{(QT)-1}$ (% $M$) further given in [8] is wrong because $T^{-1}$ % $\bar{M}$ with $T | \bar{M}$ does not exist, and neither does $Q^{-1}$ % $\bar{M}$ exist when $\gcd(Q, \bar{M}) > 1$. In the signature algorithm, it is easy to let $\gcd(Q, \bar{M}) > 1$.

Denote $x = (GW)^{-1} \delta^{\delta(\delta+1) - 1/S}$ (% $M$).

Then, the trivial solution to $x^{QT} \equiv U ((Q/H)^{1/S} \hat{G})^{-QT}$ (% $M$) does not exist when $\gcd(T, \bar{M}/T) > 1$.

Due to stipulating $T \geq 2^n$ in the key generation algorithm, the time complexity of finding out a random solution to $x^{QT} \equiv U((Q/H)^{1/S} \hat{G})^{-QT}$ is at least $\max(O(2^{n-1}), O(M/(QT)))$ through the probabilistic algorithm[10].

If a solution to $x^{QT} \equiv U((Q/H)^{1/S} \hat{G})^{-QT}$ is found through the discrete logarithm method, the probability that the solution is just equal to $(GW)^{-1} \delta^{\delta(\delta+1) - 1/S}$ (% $M$) is at most $1/2^n$.

If denote $x = ((GW)^{-1} \delta^{\delta(\delta+1) - 1/S})^T$ (% $M$), then $x^Q \equiv U ((Q/H)^{1/S} \hat{G})^{-QT}$ (% $M$).

When $\gcd(Q, \bar{M}) > 5$ and $M/Q > 2^n$, seeking a solution to $x^Q \equiv U ((Q/H)^{1/S} \hat{G})^{-QT}$ is also at least the discrete logarithm problem.

### 6.2 Forging Attack in [8] May Be Easily Avoided through Turning $D | (\delta Q - W)$ to $D | (\delta Q - WH)$

In REESSE1+[4], we definitely pointed out that $Q \neq Q_1$, where $Q$ is produced currently, and $Q_1$ is any of signature foreparts produced ever before. Of course, $Q \neq Q_1$ implied that the linear combination of $Q_1$ with $Q_2$ should be excluded from signature foreparts. However, such exclusion is infeasible in polynomial time.

Therefore, in practical applications, it is suggested as a shortcut that users move the parameter $H$ in $Q \equiv (R G_0)^S H \delta$ (% $M$) into $D | (\delta Q - W)$, and make $D | (\delta Q - W)$ become $D | (\delta Q - WH)$. In this wise, the forgery attack in [8] is easily avoided, namely $Q'$ can not be forged out at least in polynomial time.

Notice that correspondingly, the $\lambda S$ in the signature algorithm and the discriminant in the verification algorithm should also be adjusted.

## 7  Conclusion

The above rebuttal shows that each or the combination of (1″), $q_{u+1} > q_u \Delta$, and *table* 1 is not sufficient for $f(i) + f(j) = f(k)$, there exist logic errors in the deduction of (3), and *alg*.1 based on *fact* 4 and *alg*.2 based on *table* 1 are not valid. Additional, the signature forgery attack in [8] is easily avoided. Hence, the conclusion of [8] that REESSE1+ is not secure at all (which connotes that [8] can extract a related private key from any public key in REESSE1+) is completely incorrect, as long as $\Omega$ is fitly selected, REESSE1+ with $C_x \equiv A_x W^{f(x)}$ (% $M$) is secure, and the private key attack in [8] like [6] is a pseudo attack.

The authors of [8] attempt to convince people or credulous one of their opinion through an example woven elaborately, and their purpose is to want to suffocate REESSE1+, suppress us, and elevate themselves. Especially, [8] like [6] does not list the origin of idea of the continued fraction analysis of REESSE1, and falsifies the data at *table* 2, which violates scientific research ethics and honesty.

We welcome unmalicious, co-promotive, and normal academic criticism which is utterly necessary.

## Remark

The first version of this paper was sent to the authors of [8] via email on Mar. 6, 2007, and the draft of this revised version was sent to the authors of [8] via email between Oct. 23 and Nov. 12, 2009 repeatedly.

The authors of [8] revised section 5 of [8] on Mar. 12, 2007 after they read this paper and the eprint.iacr.org′s demand that [8] should be withdrawn or modified, but the modification avoided the heavy and chose the light.